\documentclass[aps,prl,preprint,tightenlines,superscriptaddress,byrevtex]{revtex4}
\usepackage{graphicx} 
\usepackage{dcolumn}  

\graphicspath{{ps}}

\begin{document}


\preprint{\vbox{ \hbox{   }
                 \hbox{BELLE-CONF-0557}
                 \hbox{LP2005-192}
                 \hbox{EPS05-532}
}}

\title{ \quad\\[0.5cm] \boldmath Hadronic Mass Moments in $B\to
  X_c\ell \nu$ Decays}

\affiliation{Aomori University, Aomori}
\affiliation{Budker Institute of Nuclear Physics, Novosibirsk}
\affiliation{Chiba University, Chiba}
\affiliation{Chonnam National University, Kwangju}
\affiliation{University of Cincinnati, Cincinnati, Ohio 45221}
\affiliation{University of Frankfurt, Frankfurt}
\affiliation{Gyeongsang National University, Chinju}
\affiliation{University of Hawaii, Honolulu, Hawaii 96822}
\affiliation{High Energy Accelerator Research Organization (KEK), Tsukuba}
\affiliation{Hiroshima Institute of Technology, Hiroshima}
\affiliation{Institute of High Energy Physics, Chinese Academy of Sciences, Beijing}
\affiliation{Institute of High Energy Physics, Vienna}
\affiliation{Institute for Theoretical and Experimental Physics, Moscow}
\affiliation{J. Stefan Institute, Ljubljana}
\affiliation{Kanagawa University, Yokohama}
\affiliation{Korea University, Seoul}
\affiliation{Kyoto University, Kyoto}
\affiliation{Kyungpook National University, Taegu}
\affiliation{Swiss Federal Institute of Technology of Lausanne, EPFL, Lausanne}
\affiliation{University of Ljubljana, Ljubljana}
\affiliation{University of Maribor, Maribor}
\affiliation{University of Melbourne, Victoria}
\affiliation{Nagoya University, Nagoya}
\affiliation{Nara Women's University, Nara}
\affiliation{National Central University, Chung-li}
\affiliation{National Kaohsiung Normal University, Kaohsiung}
\affiliation{National United University, Miao Li}
\affiliation{Department of Physics, National Taiwan University, Taipei}
\affiliation{H. Niewodniczanski Institute of Nuclear Physics, Krakow}
\affiliation{Nippon Dental University, Niigata}
\affiliation{Niigata University, Niigata}
\affiliation{Nova Gorica Polytechnic, Nova Gorica}
\affiliation{Osaka City University, Osaka}
\affiliation{Osaka University, Osaka}
\affiliation{Panjab University, Chandigarh}
\affiliation{Peking University, Beijing}
\affiliation{Princeton University, Princeton, New Jersey 08544}
\affiliation{RIKEN BNL Research Center, Upton, New York 11973}
\affiliation{Saga University, Saga}
\affiliation{University of Science and Technology of China, Hefei}
\affiliation{Seoul National University, Seoul}
\affiliation{Shinshu University, Nagano}
\affiliation{Sungkyunkwan University, Suwon}
\affiliation{University of Sydney, Sydney NSW}
\affiliation{Tata Institute of Fundamental Research, Bombay}
\affiliation{Toho University, Funabashi}
\affiliation{Tohoku Gakuin University, Tagajo}
\affiliation{Tohoku University, Sendai}
\affiliation{Department of Physics, University of Tokyo, Tokyo}
\affiliation{Tokyo Institute of Technology, Tokyo}
\affiliation{Tokyo Metropolitan University, Tokyo}
\affiliation{Tokyo University of Agriculture and Technology, Tokyo}
\affiliation{Toyama National College of Maritime Technology, Toyama}
\affiliation{University of Tsukuba, Tsukuba}
\affiliation{Utkal University, Bhubaneswer}
\affiliation{Virginia Polytechnic Institute and State University, Blacksburg, Virginia 24061}
\affiliation{Yonsei University, Seoul}
  \author{K.~Abe}\affiliation{High Energy Accelerator Research Organization (KEK), Tsukuba} 
  \author{K.~Abe}\affiliation{Tohoku Gakuin University, Tagajo} 
  \author{I.~Adachi}\affiliation{High Energy Accelerator Research Organization (KEK), Tsukuba} 
  \author{H.~Aihara}\affiliation{Department of Physics, University of Tokyo, Tokyo} 
  \author{K.~Aoki}\affiliation{Nagoya University, Nagoya} 
  \author{K.~Arinstein}\affiliation{Budker Institute of Nuclear Physics, Novosibirsk} 
  \author{Y.~Asano}\affiliation{University of Tsukuba, Tsukuba} 
  \author{T.~Aso}\affiliation{Toyama National College of Maritime Technology, Toyama} 
  \author{V.~Aulchenko}\affiliation{Budker Institute of Nuclear Physics, Novosibirsk} 
  \author{T.~Aushev}\affiliation{Institute for Theoretical and Experimental Physics, Moscow} 
% ...
\collaboration{The Belle Collaboration}

\begin{abstract}
We report measurements of the first and second moments of the hadronic
invariant mass squared distribution, $\langle M^2_X\rangle$ and
$\langle(M^2_X-\langle M^2_X\rangle)^2\rangle$, in $B\to
X_c\ell\nu$~decays for minimum lepton momenta ranging from 0.7 to
1.5~GeV/$c$ in the $B$~meson rest frame. The measurement uses $B\bar
B$~events in which the hadronic decay of one $B$~meson is fully
reconstructed and the semileptonic decay of the other $B$ is inferred
from the presence of an identified lepton. These results are obtained
from a 140~fb$^{-1}$ data sample collected near the
$\Upsilon(4S)$~resonance with the Belle detector at the KEKB
asymmetric energy $e^+e^-$~collider.
\end{abstract}


\maketitle

{\renewcommand{\thefootnote}{\fnsymbol{footnote}}}
\setcounter{footnote}{0}

Recently, there have been intense theoretical and experimental efforts
toward predicting~\cite{ref:1,ref:2,ref:3,ref:4} and
measuring~\cite{ref:5,ref:6,ref:7} the moments of the hadronic
invariant mass squared distribution in $B\to X_c\ell\nu$~decays. The
idea is that using the Operator Product Expansion (OPE)~\cite{ref:7b},
the hadronic mass moments (and other inclusive observables in
$B$~decays) can be predicted in terms of the $b$-quark mass~$m_b$ and
other non-perturbative parameters. Conversely, by measuring the
moments of $B$~decay spectra and the semileptonic $B$~decay rate, one
can then extract these parameters and $|V_{cb}|$~\cite{ref:8}.

The analysis is based on the data recorded with the Belle
detector~\cite{ref:9} at the asymmetric energy $e^+e^-$~collider
KEKB~\cite{ref:10}, operating at a center-of-mass (c.m.) energy near
the $\Upsilon(4S)$~resonance. KEKB consists of a low energy ring (LER)
of 3.5 GeV positrons and a high energy ring (HER) of 8 GeV
electrons. The Belle detector is a large-solid-angle magnetic
spectrometer consisting of a three-layer silicon vertex detector
(SVD), a 50-layer central drift chamber (CDC), an array of aerogel
threshold \v{C}erenkov counters (ACC), a barrel-like arrangement of
time-of-flight scintillation counters (TOF), and an electromagnetic
calorimeter comprised of CsI(Tl) crystals (ECL) located inside a
super-conducting solenoid coil that provides a 1.5~T magnetic
field. The responses of the ECL, CDC ($dE/dx$) and ACC detectors are
combined to provide clean electron identification. Muons are
identified in the instrumented iron flux-return (KLM) located outside
of the coil. Charged hadron identification relies on the information
from the CDC, ACC and TOF~sub-detectors.

The $\Upsilon(4S)$~dataset used for this study corresponds
to an integrated luminosity of 140~fb$^{-1}$, or about
152 million $B\bar B$~events. Another 15~fb$^{-1}$ taken 60~MeV below
the resonance are used to subtract the non-$B\bar B$ (continuum)
background. Full detector simulation based on GEANT~\cite{ref:11} is
applied to Monte Carlo (MC) simulated events. The size of the MC samples is
equivalent to about 2.4 times the integrated luminosity. At the
generator level, the decay $B\to D^*\ell\nu$ is simulated using a
HQET-based model~\cite{ref:12}. The ISGW2 model~\cite{ref:13} is used
for the decays $B\to D\ell\nu$ and $B\to D^{**}\ell\nu$. The
non-resonant $B\to D^{(*)}\pi\ell\nu$~component is generated according
to the model of Goity and Roberts~\cite{ref:14}. QED~bremsstrahlung in
semileptonic decays is simulated by the PHOTOS~package~\cite{ref:14b}.

After selecting hadronic events~\cite{ref:15}, we fully reconstruct
the hadronic decay of one $B$~meson ($B_\mathrm{tag}$) using
the decay modes $B^+\to\bar D^{(*)0}\pi^+, \bar D^{(*)0}\rho^+, \bar
D^{(*)0}a_1^+$ and $B^0\to D^{(*)-}\pi^+, D^{(*)-}\rho^+,
D^{(*)-}a_1^+$~\cite{ref:16}. Pairs of photons satisfying
$E_\gamma>50$~MeV and 117~MeV/$c^2<m(\gamma\gamma)<150$~MeV/$c^2$ are
combined to form $\pi^0$~candidates. $K^0_S$~mesons are
reconstructed from pairs of oppositely charged tracks with invariant
mass within $\pm 30$~MeV/$c^2$ of the $K^0_S$~mass and decay
vertex displaced from the interaction point. Candidate $\rho^+$ and
$\rho^0$~mesons are reconstructed in the $\pi^+\pi^0$ and $\pi^+\pi^-$
decay modes, requiring their invariant masses to be within $\pm
150$~MeV/$c^2$ of the $\rho$~mass. Candidate $a_1^+$~mesons are
obtained by combining a $\rho^0$~candidate with a charged pion and
requiring an invariant mass between 1.0 and
1.6~GeV/$c^2$. $D^0$~candidates are searched for in the $K^-\pi^+$,
$K^-\pi^+\pi^0$, $K^-\pi^+\pi^+\pi^-$, $K^0_S\pi^+\pi^-$ and
$K^0_S\pi^0$~decay modes. The $K^-\pi^+\pi^+$ and
$K^0_S\pi^+$~modes are used to reconstruct
$D^+$~mesons. Charmed mesons are selected in a window corresponding to
$\pm 3$ times the mass resolution in the respective decay
mode. $D^{*+}$~mesons are reconstructed by pairing a charmed meson
with a low momentum pion, $D^{*+}\to D^0\pi^+,D^+\pi^0$. The decay
modes~$D^{*0}\to D^0\pi^0$ and $D^{*0}\to D^0\gamma$ are used to
search for neutral charmed vector mesons.

For each $B_\mathrm{tag}$~candidate, the beam-constrained
mass~$M_\mathrm{bc}$ and the energy difference~$\Delta E$ are
calculated
\begin{equation}
  M_\mathrm{bc} = \sqrt{(E_\mathrm{beam})^2-(\vec p_B)^2}~, \quad
  \Delta E = E_B-E_\mathrm{beam}~,
\end{equation}
where $E_\mathrm{beam}$ is the beam energy in the c.m.\ system and
$\vec p_B$ and $E_B$ are the 3-momentum and the energy of the
$B_\mathrm{tag}$~candidate in the same frame, respectively. The signal
region for $B_\mathrm{tag}$ is defined by the selections
$M_\mathrm{bc}>5.27$~GeV/$c^2$ and $|\Delta E|<50$~MeV. If multiple
candidates are found in a single event, the best candidate is chosen
based on $\Delta E$ and other variables.

Semileptonic decays of the other $B$~meson ($B_\mathrm{signal}$) are
selected by searching for an identified charged lepton (electron or
muon) within the remaining particles in the event. Events with
multiple identified leptons are rejected. In the lepton momentum range
relevant to this analysis, electrons (muons) are selected with an
efficiency of 92\% (89\%) and the probability to misidentify a pion as
an electron (a muon) is 0.25\% (1.4\%)~\cite{ref:17,ref:18}. For
$B^+$~tags, we require that the lepton has a charge sign compatible
with a prompt semileptonic decay ($Q_\ell\cdot Q_B<0$, where $Q_\ell$
and $Q_B$ are the charges of the lepton and of $B_\mathrm{tag}$,
respectively). In electron events, we attempt to recover
bremsstrahlung photons by searching for a $E_\gamma<1$~GeV photon
within a 5$^\circ$~cone with respect to the electron direction. If
such a photon is found, it is merged with the reconstructed electron
candidate.

Charged and neutral particles in the event associated
neither with $B_\mathrm{tag}$ nor with the charged lepton are assigned
to the hadronic $X$~system. The missing 4-momentum in the event is
calculated, assigning the pion mass to all charged particles except
identified kaons,
\begin{equation}
  p_\mathrm{miss}=(p_\mathrm{LER}+p_\mathrm{HER})-p_{B_\mathrm{tag}}-
  p_\ell-p_X~,
\end{equation}
where the indices LER and HER refer to the colliding beams. As only
the neutrino in $B\to X_c\ell\nu$ should be missing in the event, the
missing mass is required to be consistent with zero,
$|M^2_\mathrm{miss}|<3$~GeV$^2$/$c^4$. To improve the resolution in
$M^2_X$, we constrain the neutrino mass to zero, $p_\nu=(|\vec
p_\mathrm{miss}|,\vec p_\mathrm{miss})$, and recalculate the
4-momentum of the $X$~system,
\begin{equation}
  p'_X=(p_\mathrm{LER}+p_\mathrm{HER})-p_{B_\mathrm{tag}}-p_\ell-p_\nu~.
\end{equation}
The $M^2_X$~resolution (defined as half width at the half maximum)
obtained in this way is about 800~MeV$^2/c^4$.

For the rest of the analysis, the remaining events are divided into
four sub-samples, depending on the charge of $B_\mathrm{tag}$ ($B^+$,
$B^0$) and on the lepton type (electron, muon). In each of these
sub-samples and for each lepton momentum threshold considered in
the analysis ($p^*_\ell>0.7$, 0.9, 1.1, 1.3 and
1.5~GeV/$c$~\cite{ref:18b}), the backgrounds in the
$M^2_X$~distribution are determined, taking into account contributions
from the following sources: continuum background, $B\bar B$~events with
a misreconstructed $B_\mathrm{tag}$~candidate, and background
from secondary or fake leptons. The background shapes in $M^2_X$ are
determined from the MC simulation, except for the continuum where the
off-resonance data is used. The shape of the fake muon background is
corrected by the ratio of the pion fake rate in the experimental data
over the same quantity in the MC simulation, as measured using
$K_S^0\to\pi^+\pi^-$~decays. The continuum background is scaled by the
on- to off-resonance luminosity ratio, taking into account the
cross-section difference. The combinatorial
$B_\mathrm{tag}$~background is normalized using the
$5.20$~GeV/$c^2<M_\mathrm{bc}<5.25$~GeV/$c^2$ sideband. The
normalization of the secondary or fake lepton background is found from
the real data by fitting the lepton momentum distribution. The purity
of the $B\to X_c\ell\nu$~signal depends on the sub-sample and the
lepton momentum threshold, typical values being around
75\%. Table~\ref{tab:0} shows the numbers of signal events and
purities for each combination of $B_\mathrm{tag}$~charge, lepton type
and lepton momentum threshold.
\begin{table}
  \begin{center}
    \begin{tabular}{c@{\extracolsep{.3cm}}cccc}
      \hline \hline
      \rule[-1.3ex]{0pt}{4ex}$p^*_\mathrm{min}$ & $B^+$ electron &
    $B^+$ muon & $B^0$ electron & $B^0$ muon\\
      \hline
      \rule{0pt}{2.7ex}0.7 & $3893\pm 79$ (72.3\%) & $3626\pm 84$
    (68.7\%) & $2212\pm 64$ (66.4\%) & $2154\pm 64$ (65.0\%)\\
      0.9 & $3659\pm 75$ (74.0\%) & $3484\pm 78$ (71.1\%) & $2072\pm
    58$ (66.4\%) & $2067\pm 58$ (70.2\%)\\
      1.1 & $3285\pm 70$ (75.2\%) & $3159\pm 73$ (72.9\%) & $1886\pm
    53$ (66.4\%) & $1925\pm 52$ (76.1\%)\\
      1.3 & $2742\pm 64$ (75.9\%) & $2740\pm 66$ (75.0\%) & $1595\pm
    46$ (66.4\%) & $1632\pm 47$ (79.4\%)\\
      \rule[-1.3ex]{0pt}{1.3ex}1.5 & $2152\pm 56$ (77.7\%) & $2132\pm
    56$ (76.8\%) & $1195\pm 39$ (66.4\%) & $1297\pm 41$ (83.0\%)\\
      \hline \hline
    \end{tabular}
  \end{center}
  \caption{Number of $B\to X_c\ell\nu$~signal and signal purity in the
    four sub-samples, as a function of the lepton momentum
    threshold. The yields are quoted with their statistical
    uncertainty; the corresponding signal purity is given in
    brackets.} \label{tab:0}
\end{table}

\begin{figure}
  \begin{center}
   \includegraphics{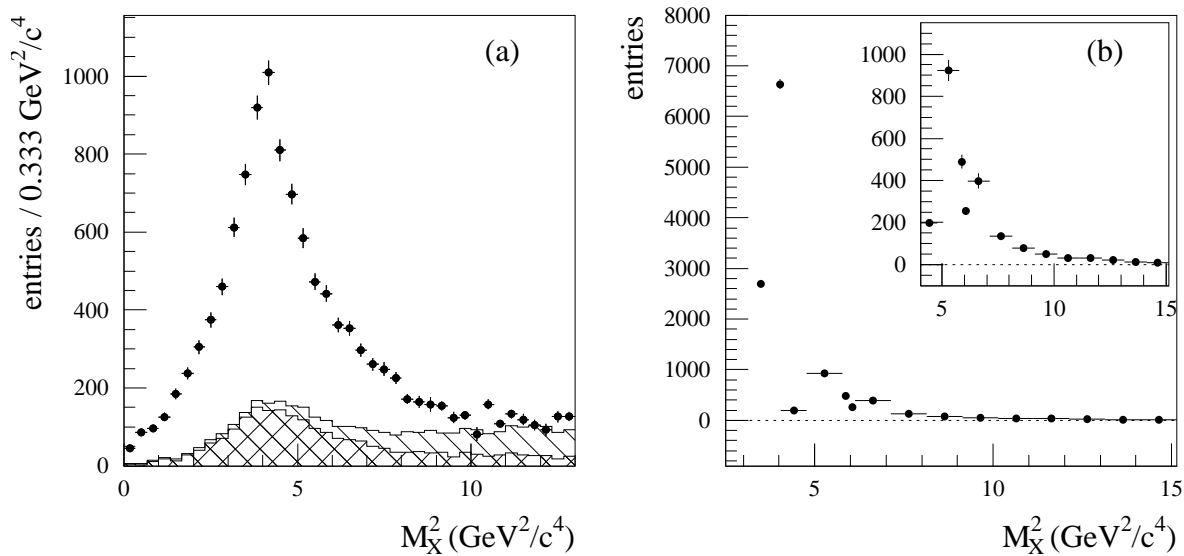}
  \end{center}
  \caption{(a) Measured and (b) unfolded $M^2_X$~distribution for
   $p^*_\ell>0.7$~GeV/$c$. On the left plot, the continuum-subtracted
   $M^2_X$~distribution is shown by points with error bars. The
   hatched histogram corresponds to the background from secondary or
   fake leptons; $B\bar B$~events in which $B_\mathrm{tag}$ is
   misreconstructed are shown by the double-hatched histogram. The
   right plot is the result of the unfolding. In both plots,
   contributions from $B^+$ and $B^0$~tags, and from electron and muon
   events are added.} \label{fig:2}
\end{figure}

In each of the four sub-samples and for each lepton momentum
threshold, the $M^2_X$~distribution is measured in 39~bins in the
range from 0 to 13 GeV$^2/c^4$ (bin width 0.333~GeV$^2/c^4$) and, after
subtraction of all backgrounds, the finite detector resolution in
$M^2_X$ is unfolded using the Singular Value Decomposition (SVD)
algorithm~\cite{ref:19}, as illustrated in Fig.~\ref{fig:2}. The
unfolded distribution has 15 bins in the range from $M^2_D$ to about
15 GeV$^2/c^4$ (bin width 1~GeV$^2/c^4$, except around the narrow
states $D$, $D^*$, $D_1$ and $D^*_2$). We calculate the first and
second moment, $\langle M^2_X\rangle$ and $\langle(M^2_X-\langle
M^2_X\rangle)^2\rangle$, for each unfolded $M^2_X$~spectrum
separately, after applying a small correction for different bin-to-bin
efficiencies. The final results for a given lepton momentum threshold
are obtained by taking the average over the four sub-samples. The
unfolding, moment calculation and averaging procedure has been studied
on MC simulated events and no significant bias has been found.

\begin{figure}
  \begin{center}
   \includegraphics{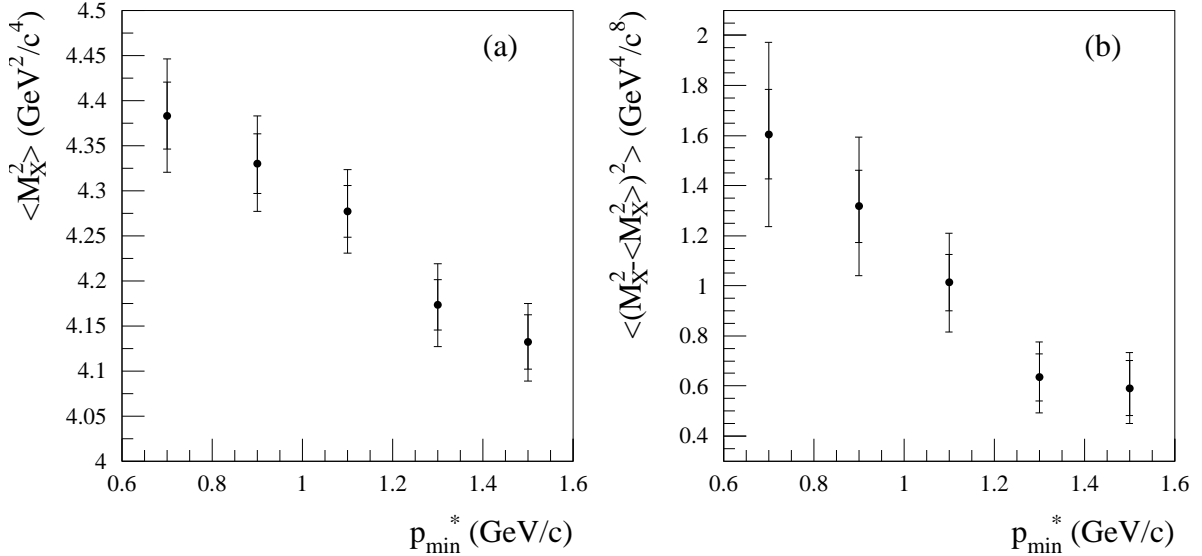}
  \end{center}
  \caption{(a) First and (b) second hadronic mass moment, $\langle
   M^2_X\rangle$ and $\langle(M^2_X-\langle M^2_X\rangle)^2\rangle$,
   for different lepton threshold momenta. The error bars indicate the
   statistical and total errors. Note that the individual moments are
   highly correlated. All results are preliminary.}
  \label{fig:3}
\end{figure}

The results for the first and second hadronic mass moment are shown in
Fig.~\ref{fig:3} and Tables~\ref{tab:1} and \ref{tab:2}. All results
are preliminary. Note that the moment measurements for different
lepton momentum thresholds are highly correlated due to overlapping
data samples. We have estimated the correlations due to this overlap
in Table~\ref{tab:3}.
\begin{table}
  \begin{center}
    \begin{tabular}{c@{\extracolsep{.3cm}}cccc}
      \hline \hline
      \rule{0pt}{2.7ex}$p^*_\mathrm{min}$ & $\langle M^2_X\rangle$ &
      detector/ & unfolding & $X_c$~model \\
      \rule[-1.3ex]{0pt}{1.3ex}(GeV/$c$) & (GeV$^2/c^4$) & background
      &	\\
            \hline
      \rule{0pt}{2.7ex}0.7 & $4.383\pm 0.037\pm 0.051$ & 0.047 & 0.004
      & 0.020\\
      0.9 & $4.330\pm 0.033\pm 0.041$ & 0.036 & 0.009 & 0.018\\
      1.1 & $4.277\pm 0.029\pm 0.035$ & 0.028 & 0.014 & 0.016\\
      1.3 & $4.173\pm 0.028\pm 0.037$ & 0.026 & 0.021 & 0.016\\
      \rule[-1.3ex]{0pt}{1.3ex}1.5 & $4.132\pm 0.030\pm 0.031$ & 0.019
      & 0.018 & 0.016\\
      \hline \hline
    \end{tabular}
  \end{center}
  \caption{First hadronic mass moment~$\langle M^2_X\rangle$ for
    different lepton threshold momenta. The first error on $\langle
    M^2_X\rangle$ is statistical and the second is the estimated
    systematic uncertainty. The right-most three columns show the
    different components of the systematic error. All results are
    preliminary.} \label{tab:1}
\end{table}
\begin{table}
  \begin{center}
    \begin{tabular}{c@{\extracolsep{.3cm}}cccc}
      \hline \hline
      \rule{0pt}{2.7ex}$p^*_\mathrm{min}$ & $\langle(M^2_X-\langle
      M^2_X\rangle)^2\rangle$ & detector/ & unfolding & $X_c$~model\\
      \rule[-1.3ex]{0pt}{1.3ex}(GeV/$c$) & (GeV$^4/c^8$) & background & \\
      \hline
      \rule{0pt}{2.7ex}0.7 & $1.605\pm 0.179\pm 0.322$ & 0.233 & 0.184
      & 0.123\\
      0.9 & $1.317\pm 0.144\pm 0.236$ & 0.174 & 0.116 & 0.110\\
      1.1 & $1.013\pm 0.113\pm 0.161$ & 0.125 & 0.047 & 0.089\\
      1.3 & $0.634\pm 0.095\pm 0.105$ & 0.079 & 0.028 & 0.064\\
      \rule[-1.3ex]{0pt}{1.3ex}1.5 & $0.591\pm 0.110\pm 0.088$ & 0.035
      & 0.068 & 0.043\\
      \hline \hline
    \end{tabular}
  \end{center}
  \caption{Same as Table~\ref{tab:1} for the second hadronic mass
  moment~$\langle(M^2_X-\langle M^2_X\rangle)^2\rangle$.} \label{tab:2}
\end{table}
\begin{table}
  \begin{center}
    \begin{tabular}{cc|ccccc|ccccc}
      \hline \hline
      \multicolumn{2}{c|}{\rule{0pt}{2.7ex}$p^*_\mathrm{min}$} &
      \multicolumn{5}{c|}{$\langle M^2_X\rangle$} &
      \multicolumn{5}{c}{$\langle(M^2_X-\langle
      M^2_X\rangle)^2\rangle$}\\
      \multicolumn{2}{c|}{\rule[-1.3ex]{0pt}{1.3ex}(GeV/$c$)} & 0.7 &
      0.9 & 1.1 & 1.3 & 1.5 & 0.7 & 0.9 & 1.1 & 1.3 & 1.5\\
      \hline
      \rule{0pt}{2.7ex} & 0.7 & 1.000 & 0.922 & 0.807 & 0.620 & 0.533
      & 0.782 & 0.720 & 0.622 & 0.471 & 0.415\\
      & 0.9 & & 1.000 & 0.875 & 0.672 & 0.579 & 0.705 & 0.781 & 0.674
      & 0.511 & 0.450\\
      $\langle M^2_X\rangle$ & 1.1 & & & 1.000 & 0.768 & 0.661 & 0.561
      & 0.622 &
      0.770 & 0.583 & 0.514\\
      & 1.3 & & & & 1.000 & 0.861 & 0.347 & 0.384 & 0.476 & 0.759 &
      0.670\\
      \rule[-1.3ex]{0pt}{1.3ex} & 1.5 & & & & & 1.000 & 0.342 & 0.379
      & 0.470 & 0.749 & 0.778\\
      \hline
      \rule{0pt}{2.7ex} & 0.7 & & & & & & 1.000 & 0.902 & 0.728 &
      0.456 & 0.440\\
      & 0.9 & & & & & & & 1.000 & 0.807 & 0.506 & 0.487\\
      $\langle(M^2_X-\langle M^2_X\rangle)^2\rangle$ & 1.1 & & & & & &
          & & 1.000 & 0.627 & 0.603\\
      & 1.3 & & & & & & & & & 1.000 & 0.963\\
      \rule[-1.3ex]{0pt}{1.3ex} & 1.5 & & & & & & & & & & 1.000\\
      \hline \hline
    \end{tabular} 
  \end{center}
  \caption{Estimated correlation coefficients for the different moment
    measurements due to overlapping data samples.} \label{tab:3}
\end{table}

We consider three sources of systematic error, shown separately in
columns three to five of Tables~\ref{tab:1} and \ref{tab:2}: the
uncertainty related to the detector modeling and the background
subtraction, the uncertainty related to unfolding and the moment
extraction procedure, and the uncertainty related to the $X_c$~model
in the MC simulation. The total systematic error is the quadratic sum
of these three components.

The uncertainty related to the detector modeling and the background
subtraction is estimated by varying the normalizations of the
different background components and some of the selections used in
the analysis (namely the $B_\mathrm{tag}$~signal region and the
requirement on $M^2_\mathrm{miss}$). The uncertainty related to
unfolding and moment extraction is obtained by varying the effective
rank parameter in the SVD~algorithm, dis- and enabling the bin-to-bin
efficiency correction and changing the binning of the unfolded
distribution. The $X_c$~model uncertainty is determined by varying the
fractions of $B\to D^*\ell\nu$, $B\to D\ell\nu$ and $B\to
D^{**}/D^{(*)}\pi\ell\nu$ within $\pm 10\%$, $\pm 10\%$ and $\pm
30\%$, respectively, and summing the individual variations in
quadrature. These ranges of variation roughly correspond to the
experimental uncertainties~\cite{ref:20} in the isospin averaged
branching ratios.

In summary, we have measured the first and second moments of the
hadronic invariant mass squared distribution, $\langle M^2_X\rangle$
and $\langle(M^2_X-\langle M^2_X\rangle)^2\rangle$, in $B\to
X_c\ell\nu$~decays for minimum lepton momenta ranging from 0.7 to
1.5~GeV/$c$ in the $B$~rest frame. The results obtained are compatible
with theoretical expectations~\cite{ref:1} and recent measurements
from other experiments~\cite{ref:6,ref:7}. In addition, we have
estimated the correlations of the moment measurements. These
measurements can be used as input to a global fit analysis, which is
expected to lead to an improved determination of the CKM matrix
element~$|V_{cb}|$.

We thank the KEKB group for the excellent operation of the
accelerator, the KEK cryogenics group for the efficient
operation of the solenoid, and the KEK computer group and
the National Institute of Informatics for valuable computing
and Super-SINET network support. We acknowledge support from
the Ministry of Education, Culture, Sports, Science, and
Technology of Japan and the Japan Society for the Promotion
of Science; the Australian Research Council and the
Australian Department of Education, Science and Training;
the National Science Foundation of China under contract
No.~10175071; the Department of Science and Technology of
India; the BK21 program of the Ministry of Education of
Korea and the CHEP SRC program of the Korea Science and
Engineering Foundation; the Polish State Committee for
Scientific Research under contract No.~2P03B 01324; the
Ministry of Science and Technology of the Russian
Federation; the Ministry of Higher Education, 
Science and Technology of the Republic of Slovenia;  
the Swiss National Science Foundation; the National Science Council and
the Ministry of Education of Taiwan; and the U.S.\
Department of Energy.


\begin{thebibliography}{99}
\bibitem{ref:16} Throughout this paper, the inclusion of the charge
  conjugate mode is implied.
\bibitem{ref:18b} Throughout this paper, quantities calculated in the
$B$~meson rest frame are denoted by an asterisk.
\end{thebibliography}
\end{document}